\input epsf.tex

\documentstyle[12pt]{article}

\newcommand{\bmat}{\left(\begin{array}}
\newcommand{\emat}{\end{array}\right)}
\def\NPB#1#2#3{Nucl. Phys. B{#1} (19#2) #3}
\def\PLB#1#2#3{Phys. Lett. B{#1} (19#2) #3}

\def\PRD#1#2#3{Phys. Rev. D{#1} (19#2) #3}

\def\yzero{\smash{\hbox{$y\kern-4pt\raise1pt\hbox{${}^\circ$}$}}}

\def\beq{\begin{equation}}
\def\eeq{\end{equation}}
\def\beqa{\begin{eqnarray}}
\def\eeqa{\end{eqnarray}}

\def\-{\hphantom{-}}
\def\ov{\overline}
\def\s2{\frac{1}{\sqrt2}}

\def\beq{\begin{equation}}
\def\eeq{\end{equation}}
\def\beqa{\begin{eqnarray}}
\def\eeqa{\end{eqnarray}}

\def\Tr{{\rm Tr \,}}
\def\diag{{\rm diag \,}}
\def\IF{\relax{\rm I\kern-.18em F}}
\def\II{\relax{\rm I\kern-.18em I}}
\def\IP{\relax{\rm I\kern-.18em P}}
\def\IC{\relax\hbox{\kern.25em$\inbar\kern-.3em{\rm C}$}}

\def\cp{{\cal P}}
\def\IC{\bf C}
\def\IR{\bf R}
\def\IS{\bf S}
\def\IZ{\bf Z}
\def\IT{\bf T}

\def\id{{\bf 1}}

\def\NN{{\cal N}}
\def\Dsl{\,\raise.15ex\hbox{/}\mkern-13.5mu D} 
 \def\cp#1{\relax\ifmmode {\IP\kern-2pt{}_{#1}}\else $\IP\kern-2pt{}_{#1}$\=fi}

\begin{document}

\makeatletter \@addtoreset{equation}{section} \makeatother
\renewcommand{\theequation}{\thesection.\arabic{equation}}
\pagestyle{empty}
\rightline{CAB-IB 2919500, CERN-TH/2000-320, CTP-MIT-3042,
FTUAM-00/23, IFT-UAM/CSIC-00-37}
\rightline{\tt hep-ph/0011132}
\vspace{0.5cm}
\begin{center}
\LARGE{\bf
 Intersecting Brane Worlds
\\[10mm]}
\medskip
\large{G.~Aldazabal$^{1}$, S. Franco$^2$,
L.~E.~Ib\'a\~nez$^3$, R. Rabad\'an$^3$, A. M. Uranga$^4$
\\[2mm]}
\small{$^1$ Instituto Balseiro, CNEA and CONICET,\\[-0.3em]
 Centro At\'omico Bariloche, 8400 S.C. de Bariloche, Argentina.\\[1mm]
$^2$ Center for Theoretical Physics,\\[-0.3em]
Massachusetts Institute of Technology,
Cambridge MA 02139, U.S.A.\\[1mm]
$^3$ Departamento de F\'{\i}sica Te\'orica C-XI
and Instituto de F\'{\i}sica Te\'orica  C-XVI,\\[-0.3em]
Universidad Aut\'onoma de Madrid,
Cantoblanco, 28049 Madrid, Spain.\\[1mm]
$^4$ Theory Division, CERN, Geneva, Switzerland.
\\[3mm]}

\small{\bf Abstract} \\[3mm]
\end{center}

{\small
\begin{center}
\begin{minipage}[h]{14.5cm}
It is  known that chiral fermions naturally appear at certain
intersections of branes at angles. Motivated by this fact, we propose a
string scenario in which different standard model gauge interactions
propagate on different (intersecting) brane worlds, partially wrapped in
the extra dimensions. Quarks and leptons live at brane intersections,
and are thus located at different positions in the extra dimensions.
Replication of families follows naturally from the fact that the branes
generically intersect at several points. Gauge and Yukawa couplings can be
computed in terms of the compactification radii. Hierarchical Yukawa
couplings appear naturally, since amplitudes involving three different
intersections are proportional to $e^{-A_{ijk}}$, where $A_{ijk}$ is the
area of a string world-sheet extending among the intersections. The models are
non-supersymmetric but the string scale may be lowered down to 1-10 TeV.
The proton is however stable due to a set of discrete symmetries arising
from world-sheet selection rules, exact to all orders in perturbation
theory. The scenario has some distinctive features like the presence of
KK, winding and other new excited states (`gonions'), with masses below
the string scale and accessible to accelerators. The models contain scalar
tachyons with the quantum numbers of standard $SU(2)\times U(1)$ Higgs
doublets, and we propose that they induce electroweak symmetry breaking in
a somewhat novel way. Specific string models with D4-branes wrapping on
$\IT^2\times (\IT^2)^2/\IZ_N$, leading to three-family semirealistic
spectra, are presented in which the above properties are exemplified.
\end{minipage}
\end{center}}
\newpage
\setcounter{page}{1} \pagestyle{plain}
\renewcommand{\thefootnote}{\arabic{footnote}}
\setcounter{footnote}{0}

\section{Introduction}

Two of the most important aspects of the observed fermion spectrum of the
standard model (SM) are its chirality and the family replication. Any
fundamental theory explaining the structure of the SM should thus give an
understanding of these two very prominent features. With the developments
of string theory of the last five years we have learnt that a natural setting
to understand  gauge interactions in this context is that of Type II
D$p$-branes, which contain gauge fields localized in their world-volume.
However, D$p$-branes isolated on a smooth space have extended supersymmetry,
and hence do not lead to chiral fermions. Thus, for example, Type IIB
D3-branes at a smooth point in transverse space have $\NN=4$ supersymmetry
on their four-dimensional world-volume.

A simple possibility to obtain chirality is to locate the
D3-branes on some singularity in transverse space, the simplest
possibility being a $\IC^3/\IZ_N$ orbifold singularity \break
\cite{dm,dgm}\footnote{Specific semirealistic string models based
on this possibility with the gauge group of the SM or a
left-right extension were recently constructed  in \cite{aiqu}.
See e.g. \cite{evenmore} for other attempts to build realistic 
string models of the brane world scenario \cite{aadd,otherbw,aab}}.
There is however an interesting alternative to obtain chiral
fermions, which has not being very much exploited in the past
from the phenomenological viewpoint. As first pointed out in
\cite{bdl}, when D$p$-branes intersect at non-vanishing angles,
open string stretched between them may give rise to chiral
fermions living at the intersection. Our purpose in the present
article is to study the phenomenological potential of this kind
of configurations, in which the observed quarks and leptons are
associated to intersections among D$p$-branes. In our setting the
different SM gauge interactions propagate on different branes,
and chiral fermions propagate at their intersections. That is, we
have gauge bosons propagating on {\it intersecting brane worlds},
with quarks and leptons populating the intersections.

\begin{figure}
\begin{center}
\centering
\epsfysize=10cm
\leavevmode
\epsfbox{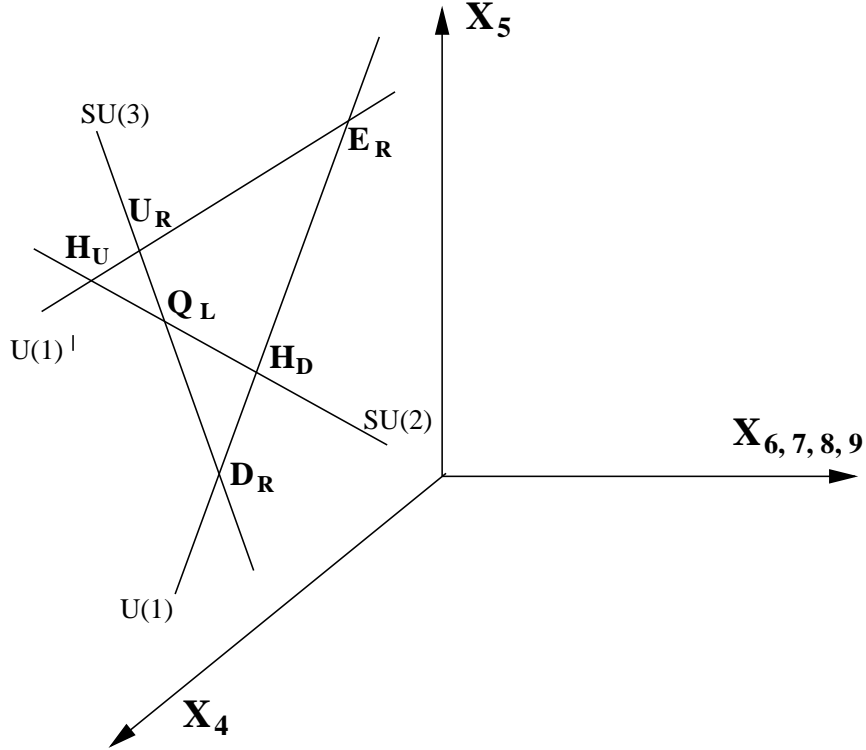}
\end{center}
\caption[]{\small A simplified picture of the intersecting brane
world scenario. Each gauge interaction propagates along a D-brane
with four flat dimensions (not shown in the figure), and
partially wrapped on a cycle in the internal space parameterized
by $X_4$, $X_5$ (a two-torus in our models). All branes are
transverse to the space parameterized by $X_6$, $X_7$, $X_8$,
$X_9$. Chiral fermions, such as quarks and leptons, are localized
at the intersections of the wrapped branes (for simplicity, we
have shown only one such intersection, even though generically
multiple intersection points exist).} \label{oxford1}
\end{figure}

Explicit string theory compactifications with branes intersecting
at angles have appeared in \cite{bgkl}, and more extensively in
\cite{afiru}. We will concentrate in this paper on the simplest
non-trivial case, corresponding to D4-branes with one of their
world-volume dimensions wrapped on a circle inside a two-torus
\cite{afiru}. Thus the model contains different stacks of
D4-branes for the different SM gauge groups, wrapping on the
two-torus, and intersecting on four-dimensional subspaces, on
which chiral fermions propagate. For example, left-handed quarks
appear at the intersection of the $SU(3)$ D4-branes with the
$SU(2)_L$ D4-branes. A pictorial depiction of this type of
configuration is shown in Fig~\ref{oxford1}. Interestingly
enough, two non-parallel D4-branes on a torus typically intersect
at more than one point, leading to several copies of the same
matter content. Thus {\it replication of quark-lepton generations
is a generic property} in this kind of configurations. In
particular, it is easy to construct models with a triplication of
generations.

Another interesting feature of these constructions is the structure of
Yukawa couplings. Some intersection give rise also to scalar fields, which
may transform with the quantum numbers of Weinberg-Salam Higgs doublets.
Their Yukawa couplings with left(right)-handed fermions $F_L(F_R)$ will be
proportional to $\exp(-A_{ijk})$, where $A_{ijk}$ is the area of the
worldsheet extending among the intersections where the Higgs, $F_L$ and $F_R$
live. Due to this fact, it is easy to obtain a  hierarchical structure
of quark and lepton masses, as we show in some specific models.

The models we are describing are generically non-supersymmetric. In order
to avoid the gauge hierarchy problem, one may lower the string scale 
down to 1-10 TeV in the usual way \cite{aadd}, by having some or all 
of the four extra dimensions transverse to the two-torus large enough
\footnote{See \cite{ignatios} for early proposal of large volume
(heterotic) compactifications, and \cite{lykken} for an early proposal of
a low string scale.}. An important
property of these models is that they do not predict gauge coupling
unification at the string scale. Rather, the gauge coupling of each gauge
group is inversely proportional to the length of the wrapped cycle. The
values of the coupling constants can therefore be computed in terms of the
compact radii, leading to results which may be made compatible with the
experimental values. We also show that a generic difficulty in models with
a low string scale, proton stability, is naturally solved in these
configurations, where quarks and leptons live on intersecting branes. The
reason is that a proton decay process requires an overall interaction with
three incoming $SU(3)$ triplets (and no outgoing ones). Such processes
require worldsheets with an odd number of quark insertions, which do not
exist (to any order in perturbation theory).

The scenario we propose has additional specific features. We show that
there exist Kaluza-Klein (KK) and/or winding excitations of the SM gauge
bosons, which may have masses well below the string scale. Moreover there
is a new class of extra excited modes of fields at intersections (with
spin=1/2,0,1). They correspond to excited open strings stretching in the
vicinity of the intersections of the branes at angles. Their masses
are proportional to the brane angles, hence we refer to them as `gonions'.
They may have masses just above the weak scale, and thus could provide the
first signatures of a low-scale string theory.

To show that the properties advertised above are indeed possible within
the context of string theory, we construct a class of specific string
compactifications yielding the above general structure. In particular one
can easily construct a large set of three-generation models based on
D4-branes with one dimension wrapped on circles in $\IT^2\times
(\IT^4/\IZ_3)$ \cite{afiru}. They are non-supersymmetric, and typically
involve extra heavy leptons beyond those in the SM. In these specific
examples, in addition to the quarks and leptons, some intersections also
contain scalar tachyons. They are a reflection of the absence of
supersymmetry in the configuration, and signal an instability against
the rearrangement of the D4-branes, which tend to align parallel.
Interestingly enough, in some cases these tachyons have the quantum
numbers of Higgs fields, and we propose that their
presence just signals electroweak symmetry breaking.

\section{Intersecting Standard Model brane-worlds}

In order to explore the building of models with quarks and leptons at
brane intersections, we are going to consider the simplest case of sets of
D4-branes wrapping different circles on a two-torus. More specifically,
we consider the compactification of Type IIA string theory on a compact
variety of the form  $\IT^2\times {\bf B}_4$, where ${\bf B}_4$ is a
four-dimensional variety whose specific form is not necessary for the
moment \footnote{More generally, one can consider Type IIA compactified
on a six-dimensional variety (e.g. a CY manifold), which is a torus bundle
over a base ${\bf B}_4$. That is, for any small patch $U$ in ${\bf B}_4$
the local geometry factorizes as ${\IT^2}\times U$, but the global topology
is not $\IT^2\times {\bf B}_4$.}. We will skip the more technical details
here and postpone issues like tadpole cancellation and the form of the
variety ${\bf B}_4$ to section 7. We do this to simplify the presentation,
but also because the main physical issues we are discussing are present in
other more complicated string constructions with intersecting branes
\cite{bgkl,afiru}. Thus, we consider several sets of D4-branes with one
world-volume dimension wrapped on different circles within a two-torus.
Consider first a square two-torus, obtained by quotienting two-dimensional
flat space $\IR^2$ by the lattice of translations generated by the two
vectors $e_1=(1,0)$, $e_2=(0,1)$.
Thus one makes the identification $X=X+l2\pi e_1+p2\pi e_2$, $l,p\in {\IZ}$.
 The corresponding two circles are taken
with arbitrary radii $R_1$ and $R_2$, respectively. We denote by $(n,m)$ a
non-trivial cycle winding $n$ times around the cycle defined by $e_1$ and
$m$ times around the cycle defined by $e_2$. Different stacks of D4-branes
wrap around different $(n,m)$ cycles.

Consider now a stack of $N_i$ overlapping D4-branes with wrapping numbers
$(n_i,m_i)$ and a second stack of $N_j$ D4-branes with wrapping numbers
$(n_j,m_j)$. As is well known, each set of branes gives rise to a unitary
gauge factor, giving a gauge group $U(N_i)\times U(N_j)$. Notice that
these  gauge interactions live in Minkowski space plus one extra bulk
dimension, which is different for each gauge factor. Matter multiplets
arise at the intersections between the two sets of D4-branes. The number
of intersections in the two-torus is given by
\beq
I_{ij} \ =\ n_im_j\ -\ n_jm_i
\label{inter}
\eeq
At those intersections there arise chiral fermions
\footnote{Actually, in order for the fermions at the intersection
to be chiral, the transverse variety ${\bf B}_4$ mentioned above
has to fulfill certain conditions, namely it must be singular, as
we describe in Section 7. We assume in this section that this is
the case.} which transform in the   bi- \break fundamental representation
$(N_i,{\overline N}_j)$ of $U(N_i)\times U(N_j)$. These
bi-fundamentals correspond to open strings stretching between
both stacks of branes, and hence localized near the
intersections. Thus {\it chiral fermions are localized in the six
compact dimensions}. Due to the multiple number of intersections,
we obtain $I_{ij}$ copies of such fermion content
\footnote{Actually (\ref{inter}) gives the intersection number
counted with orientation, which agrees with the naive
intersection number up to a sign. A negative $I_{ij}$ indicates
that the intersections give rise to $-I_{ij}$ fermions of
opposite chirality.}, hence replication of the spectrum is a
generic feature in this type of construction. In fact, it is
quite easy to obtain configurations with three generations. To
see that, let us discuss the following example

\medskip
\noindent
 {\bf Example 1}

We choose a configuration of D4-branes at angles leading to a
left-right symmetric model. With that purpose, we consider four
sets of branes with $N_1=3$, $N_2=2$, $N_3=2$ and $N_4=1$, and
wrapping numbers
\beq
N_1:(1,0) \;\; ;\;\; N_2:(1,3) \;\; ; \;\; N_3:(1,-3) \;\; ;\;\; N_4:(-1,0).
\label{lr1}
\eeq
The resulting gauge group is $U(3)\times U(2)_L\times U(2)_R\times U(1)$.
The intersection numbers (\ref{inter}) computed using the wrappings
(\ref{lr1}), give rise to
a set of chiral fermions transforming under the non-abelian factors as
\beq
3(3,2,1)\ +\ 3({\bar 3},1,2)\ +\  3(1,2,1) \ +\ 3(1,1,2)\ +\  6(1,2,2)
\label{splr1}
\eeq
Notice that the fermion content is that of three generations of quark and
leptons. In addition there are ``Higgsino-like'' fermions transforming in
$(1,2,2)$.

The model contains four $U(1)$ gauge symmetries, from the $U(N_i)$ factors
in the different sets of branes, with generators $Q_i$, $i=1,\ldots,4$. In
fact, all fields in the model are neutral under the diagonal combination
$Q_{diag}=\sum_i Q_i$, which therefore decouples. Moreover some of the
remaining $U(1)$ symmetries are anomalous (with anomaly cancelled by a
generalized Green-Schwarz mechanism). Their detailed discussion \cite{afiru}
requires an explicit construction within string theory, to be performed in
Section~7. For our purposes here, the main conclusion from the analysis is
that the anomalous $U(1)$'s gain a mass of the order of string scale, and
that one of the surviving anomaly-free linear combinations can be identified
with the standard (B-L) symmetry of left-right symmetric models
(see section 7).

\begin{figure}
\begin{center}
\centering
\epsfysize=11cm
\leavevmode
\epsfbox{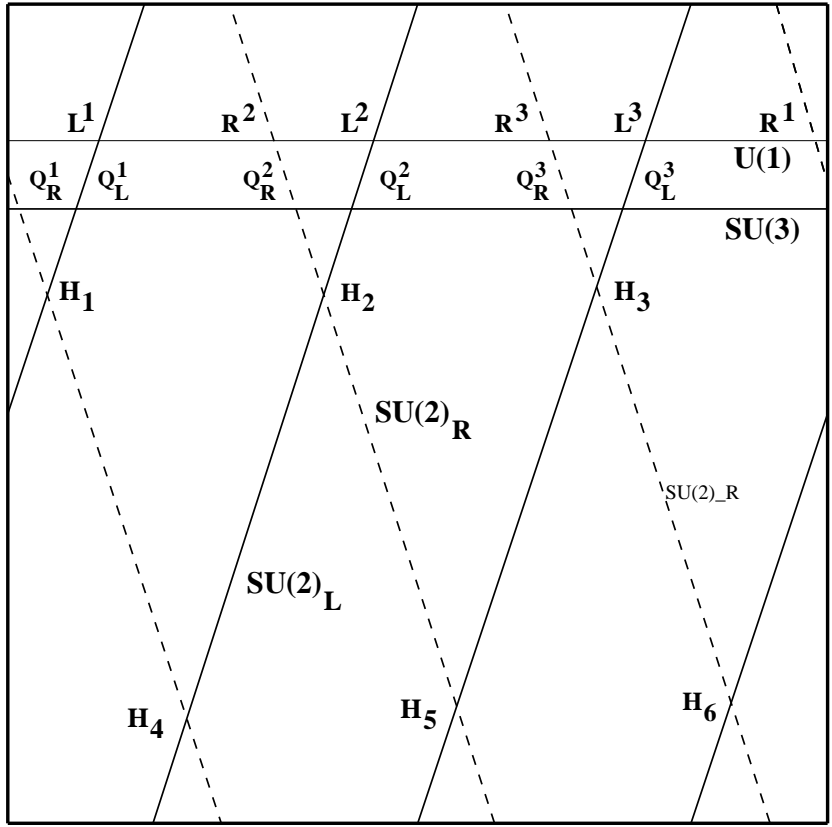}
\end{center}
\caption[]{\small D4-branes wrapping on a two-torus yielding a
three-generation $SU(3)\times SU(2)_L\times SU(2)_R\times U(1)$ model,
example 1. Gauge bosons propagate along one world-volume internal dimension,
depicted as lines. Quarks and leptons, appearing in three copies, are
located at the intersection points of different pairs of branes.}
\label{phen1}
\end{figure}

This D4-brane configuration is depicted in Fig.~\ref{phen1}. In that
figure opposite sides of the square are identified to recover the topology
of a two-torus. Gauge fields are localized along the straight lines within
the square, which represent the wrapped D4-branes. For example, the
$SU(2)_L$ branes are wrapping  three times around $e_2$ and once
around $e_1$. Chiral fermions are localized at the intersection points of the
different lines, and transform as bi-fundamental representations under the
gauge symmetries on the corresponding branes.  Notice the important
point that, since intersections take place at different points in the
two-torus, the different quarks and leptons sit at distant locations in
the extra dimensions. This turns out to be important when studying the
structure of Yukawa couplings in this kind of models (see Section 4).

\medskip
\noindent
 {\bf Example 2}

\begin{figure}
\begin{center}
\centering
\epsfysize=12cm
\leavevmode
\epsfbox{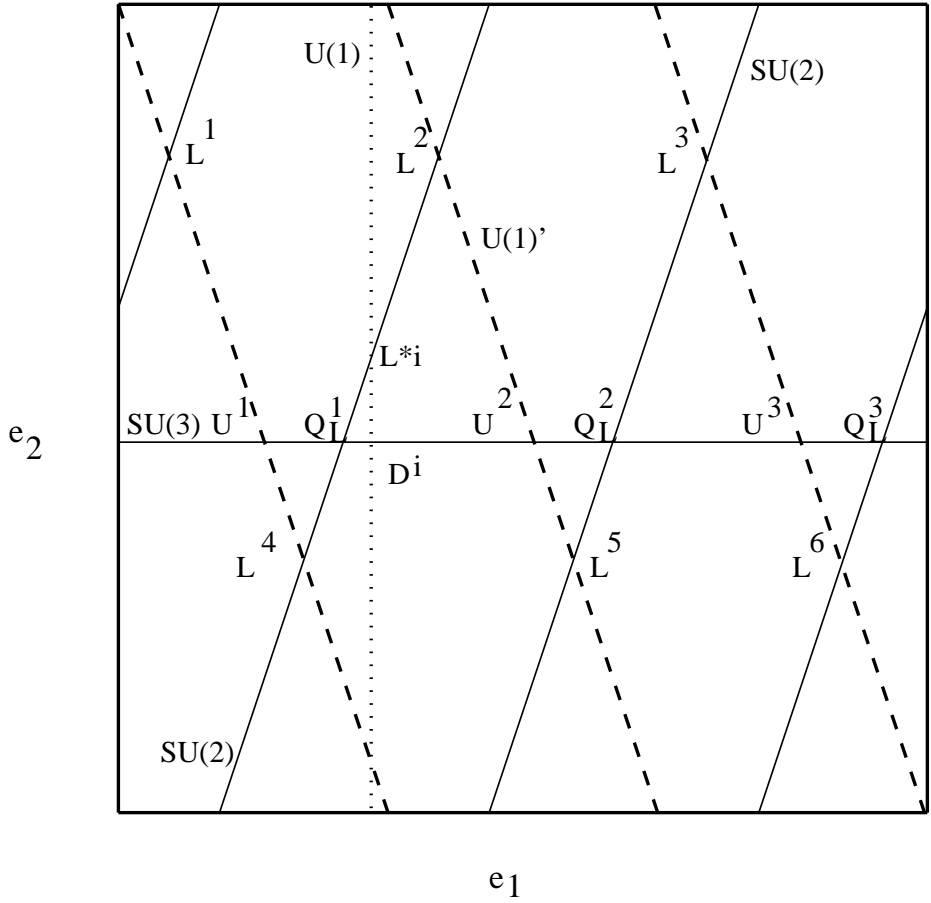}
\end{center}
\caption[]{\small D4-branes wrapping on a torus yielding a three-generation
standard model, example 2. Gauge bosons propagate along the lines, which
indicate the wrapped D4-brane world-volumes. Quarks and leptons are however
localized at the intersection points among the different branes. The
vertical $U(1)$ brane is wrapped three times along the depicted cycles,
hence leads to three (coincident) intersections with each of the remaining
branes.}
\label{phen3}
\end{figure}

There is in fact a wealth of possibilities \footnote{One can in
fact classify different families of models (wrapping numbers)
leading to three generations. See section 7.}. 
 For instance, we can
construct a Standard Model configuration, based on four sets of branes
with $N_1=3$, $N_2=2$ , $N_3=N_4=1$, and wrapping numbers
\beq
N_1:(1,0) \;\; ;\;\; N_2:(1,3) \;\; ; \;\; N_3:(0,-3) \;\; ;\;\;
N_4:(1,-3)\ .
\label{sm1}
\eeq
The intersection numbers (\ref{inter}) corresponding to these wrapping
numbers are $\pm 3$, or $\pm 6$. The resulting chiral fermions transform
as
\beq
3(3,2)\ +\ 3({\bar 3},1)\ +\ 3({\bar 3},1) \ +\
  3(1,2) \ +\ 3(1,1)  \  +\ 6(1,2)
\label{splr2 }
\eeq
under $SU(3)\times SU(2)_L$.
This  correspond to three quark-lepton generations plus an extra set of
three vector-like leptons (``Higgsinos''). This D4-brane configuration is
depicted in Fig.~\ref{phen3}. 
Concerning $U(1)$'s, again the diagonal combination decouples but
there is however an anomaly-free combination,
roughly of the form
\beqa
Q_{Y}\ & =&\ -{1\over 3}Q_1 -{1\over 2}Q_2 -Q_4\ ,
\label{bhypertoy}
\eeqa
which can be identified with standard hypercharge
\footnote{There are in this particular model additional 
anomalous and anomaly-free $U(1)$'s whose discussion is 
postponed to section 7.}. Here $Q_i$ is the
$U(1)$
generator of the $i^{th}$ stack of D4-branes.

Up to this point, we have not mentioned whether there are scalar fields at
the intersections. In general there are such fields, as we describe in
Section~7. Their existence depends on the geometry of the transverse compact
space ${\bf B}_4$. Phenomenological models require the existence of Higgs
scalars, which in our models should arise at the intersections of the
$SU(2)_L$ branes with some $U(1)$ (or $SU(2)_R$) branes. This is certainly
the case in many explicit string theory models, as we discuss in Section~7.
Leaving their detailed study for later sections, we proceed, assuming for
the time being that the models under study indeed contain appropriate
scalars to play the role of standard model Higgs fields.

\section{The gauge coupling constants}

Unlike what happens in other string scenarios, the couplings for the
different gauge factors in the model do not have the same value at the
string scale, so there is no unification of gauge couplings \footnote{The
question of gauge couplings in multiple brane scenarios has also been
considered in \cite{st,imr,akt,aiqu}.}. The gauge fields on different sets
of wrapping D4-branes have different gauge couplings $g_i$, with fine
structure constant inversely proportional to the length of the wrapped cycle
\beq
{{4\pi ^2}\over {g_i^2} } \ =\
{{M_s}\over {\lambda_{II}}} \ |(n_i,m_i)|
\label{coup}
\eeq
where $M_s$ is the string scale, $\lambda_{II}$ is the Type II string
coupling, and $|(n,m)|$ is the length of the cycle $(n,m)$.
Here we will consider the case of a general metric for the torus.
 This length
depends on the compactification radii $R_1$, $R_2$, and the angle $\theta$
between the two vectors defining the torus lattice. Distances  on a flat
torus can be seen as a scalar product of vectors with the metric
\beq
g \ = \
\left( \begin{array}{cc}
g_{11} & g_{12} \\
g_{21} & g_{22}
\end{array} \right)
 \ = \ (2\pi)^2
\left( \begin{array}{cc}
R_1^2 & R_1 R_2 \cos{\theta} \\
R_1 R_2 \cos{\theta} & R_2^2
\end{array} \right)
\label{metric}
\eeq
The length of a cycle $v=(n,m)$ is
\beqa
|(n,m)|=(g_{ab}v^a v^b)^{1/2} =2\pi\, \sqrt{n^2 R_1^2 + m^2 R_2^2 + 2 nm
R_1 R_2 \cos\theta}
\label{length}
\eeqa
Thus the relative size of the different coupling constants is governed by
the wrapping numbers $(n_i,m_i)$, the compactification radii $R_1,R_2$
and $\cos{\theta}$. In the case of an anomaly free $U(1)$ defined by a
linear combination
\beq
Q\ =\ \sum_i \ c_i\; Q_i
\eeq
the corresponding coupling is given by
\beq
{1\over {g_{U(1)}^2} } \ =\ \sum_i \ c_i\ {1\over {g_i^2}} .
\eeq
In the case of models analogous to that of example 2, one finds
\beqa
{\alpha_{QCD}}^{-1}\ & = & \ {1\over {\pi \lambda_{II}}} |(n_1,m_1)| \\
\alpha_2^{-1}  \ &  =  & \    {1\over {\pi \lambda_{II}}} |(n_2,m_2)| \\
\alpha_Y^{-1} \ & = & {(3\alpha_{QCD})}^{-1}+{(2\alpha_2)}^{-1}
                    + {1\over {\pi \lambda_{II}}} |(n_4,m_4)|
\label{coupsm}
\eeqa
where lengths are measured in string units. This leads to a weak angle
\beq
\sin^2\theta_W\ =\ \frac{g_y^2}{g_y^2 + g_2^2} \ =\ {6\over {(9+2\xi_1
+6\xi_4)}}
\label{sinw}
\eeq
where $\xi_1=g_2^2/g_1^2$ and $\xi_4=g_2^2/g_4^2$.

These are the values of the couplings at the string scale, which, since the
models are non-supersymmetric, should be of the order of 1-10 TeV to avoid
a hierarchy problem. In order to compare the values (\ref{coupsm}) with
low-energy data, running from the string scale to the weak scale should be
taken into account. The details of this running depend on the precise
low-energy content of the model \footnote{As studied in Section~6, there may
be KK/winding and other type of excitations in the region between $M_Z$
and $M_s$. They may lead to important modifications of the coupling
running to some extent analogous to those in \cite{ddg}.}. There seems to
be enough freedom in this class of models to accommodate the experimental
values by appropriately varying the choice of $(n_i,m_i)$, the radii
$R_{1,2}$, and the angle $\theta$. A detailed analysis of possibilities is
beyond the scope of this paper. For illustration, an estimation  of the
coupling constants values is performed in Section 7 for an explicit string
SM example.

\section{The structure of Yukawa couplings }

As we have seen  in previous sections, quarks, leptons and  Higgs fields
live in general at different intersections. Yukawa couplings among the
Higgs $H^i$ and two fermion states $F_R^j$, $F_L^k$ arise from a string
worldsheet stretching among the three D4-branes which cross at those
intersections. The worldsheet has a triangular shape, with vertices on the
relevant intersections, and sides within the D4-brane world-volumes. The
area of such world-sheet depends on the relative locations of the relevant
fields, and some couplings may even require world-sheets wrapped around
some direction in the two-torus.

The size of the Yukawa coupling is, for a square torus, of order
\footnote{For a general metric one just has to replace $R_1R_2 \rightarrow
R_1R_2|sin\theta| $.}
\beq
Y_{ijk}\ =\  \exp(- {\frac{R_1R_2}{\alpha '}}A_{ijk})
\label{yuk}
\eeq
where $A_{ijk}$ is the adimensional area (the torus area has been scaled
out) of the worldsheet connecting the three vertices. Since the areas
involved are typically order one in string units, corrections due to
fluctuations of the worldsheet may be important, but we expect the
qualitative behaviour to be controlled by (\ref{yuk}). This structure
makes very natural the appearance of hierarchies in Yukawa couplings of
different fermions, with a pattern controlled by the radii and the size
of the triangles.

\begin{figure}
\begin{center}
\centering
\epsfysize=14cm
\leavevmode
\epsfbox{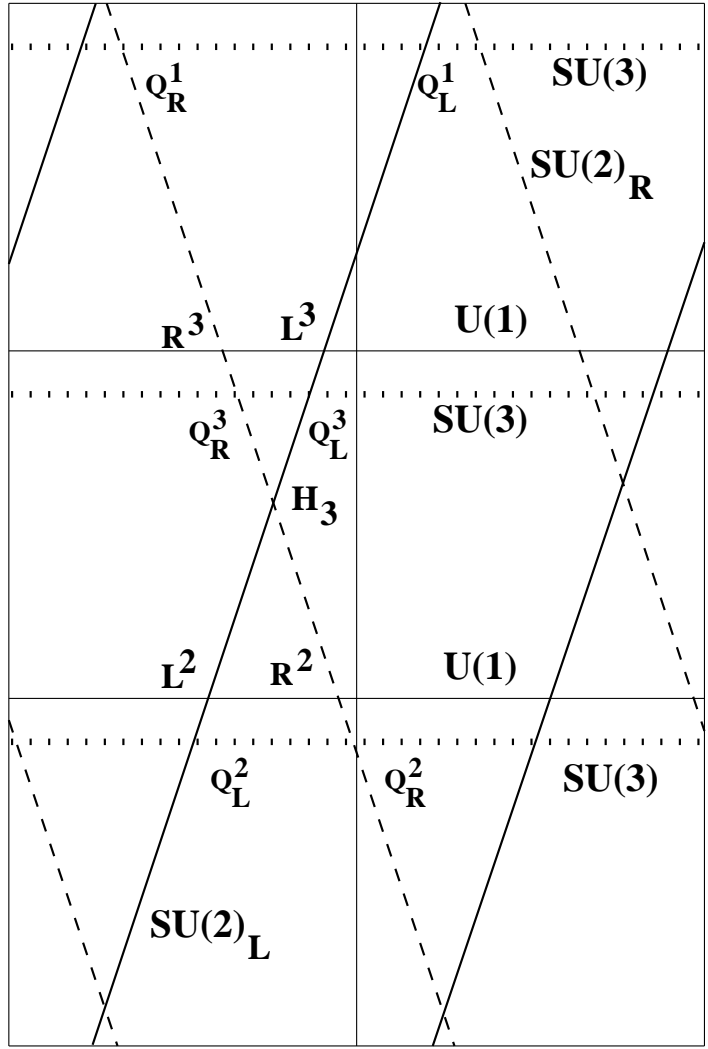}
\end{center}
\caption[]{\small The  $SU(3)\times SU(2)_L\times SU(2)_R\times U(1)$
model of Fig.~\ref{phen1}. Several torus fundamental domains are shown to
highlight the relative size of the different Yukawa couplings. To avoid
clutter, we do not show all the copies of the  branes. Also, we
only highlight the Yukawa couplings involving the Higgs field $H_3$, hence
do not show other fields living at the relevant intersections. World-sheets
giving rise to quark (lepton) Yukawa couplings correspond to triangles
with one vertex ($H_3$) containing the Higgs and other two vertices
$Q_L^i,Q_R^i$ ($L^i,R^i$) containing the quarks (leptons).}
\label{inter17}
\end{figure}

The cycle wrapped by the $i^{th}$ D4-brane around a rectangular torus is
given by a straight line equation
\beq
X_2^i\ =\ a_i (2\pi R_2)\ +\ {{m_i R_2}\over {n_i R_1}}\ X_1^i,
\label{str}
\eeq
and the $i^{th}$ and $j^{th}$ D4-branes intersect at the point:
\beq
(X_1,X_2)_{ij} \ =\  {{2\pi}\over {I_{ij}}}\
(n_in_j(a_i-a_j)R_1, (a_in_im_j-a_jn_jm_i) R_2)
\label{punto}
\eeq
where $I_{ij}$ is the intersection number for the two D-branes. Hence, the
area of each triangle depends not only on the wrapping numbers $(n_i,m_i)$
but also on the $a_i$'s.

It is clear from the above structure that one can easily generate hierarchies
of Yukawa couplings and possibly interesting textures for suitable choices
of the free parameters in the models, i.e. the wrapping numbers, the compact
radii (and the angle between axes for non-square tori), and the parameters
$a_i$ of each stack of branes. A systematic search for phenomenologically
interesting textures is beyond the scope of this paper. However, let us
illustrate the idea by considering as an example the left-right symmetric
model considered in section 2 (example 1).

The configuration is shown for the case of a square lattice
in Fig.~\ref{inter17}, where in order to get
a better visualization, we include several fundamental domains of the torus.
The left(right)-handed quarks are denoted by $Q_L^i(Q_R^i)$ and the
left(right)-handed leptons by $L^i(R^i)$. Scalars transforming as $(1,2,2)$
appear at the intersection of the $SU(2)_L$ and $SU(2)_R$ branes and are
denoted by $H_i$. Let us first consider the structure of quark Yukawa
couplings to one of the Higgs fields, say $H_3$. For the choice of brane
positions shown in Fig.~\ref{inter17}, the couplings of $H_3$ to the three
generations of quarks
\beqa
h_3 H_3 Q_L^3 Q_R^3 \quad ; \quad
h_2 H_3 Q_L^2 Q_R^2 \quad ; \quad
h_1 H_3 Q_L^1 Q_R^1
\eeqa
with $h_3>>h_2>>h_1$ for sufficiently large radii. In particular,
notice that the example in the figure would give rise to a
hierarchy (before QCD loop corrections) 
$m_b>m_{\tau}>m_{\mu}>m_s>m_d>m_e$  and $m_t>m_c>m_u$, corresponding to the
relative sizes of the triangles.

In this example there are additional Yukawa couplings, which are perhaps
more evident in the representation in Fig.~\ref{phen1}, involving other 
Higgs fields like $H_4$, $H_5$ and $H_6$. In particular one has 
additional couplings of
the form
\beqa
h'H_4Q_L^1Q_R^2\ &+&\ h''H_4Q_L^2Q_R^1 +\ \nonumber \\
h'H_5Q_L^2Q_R^3\ &+&\ h''H_5Q_L^3Q_R^2 +\ \nonumber \\
h'H_6Q_L^3Q_R^1\ &+&\ h''H_6Q_L^1Q_R^3
\label{yukmix}
\eeqa
 Hence, assuming $H_3$
has the dominant vev as above, vevs for $H_4$, $H_5$ 
or $H_6$ would contribute to
non-diagonal entries in the quark mass matrix,
giving rise to generation mixing. Clearly, a similar pattern
holds for leptons.

In fact, the existence of mixing is generic in this class of brane models.
This is explicit also in the SM example of Figure~\ref{phen3}. If we assume
that the Higgs fields which couple to the u-type quarks arise at the
intersections labeled $L^i$, it is clear from the figure that the scalars
in the locations $L^4,L^5,L^6$ couple diagonally to the quarks whereas
those in $L^1,L^2,L^3$ generate off-diagonal couplings.

\begin{figure}
\begin{center}
\centering
\epsfysize=16cm
\leavevmode
\epsfbox{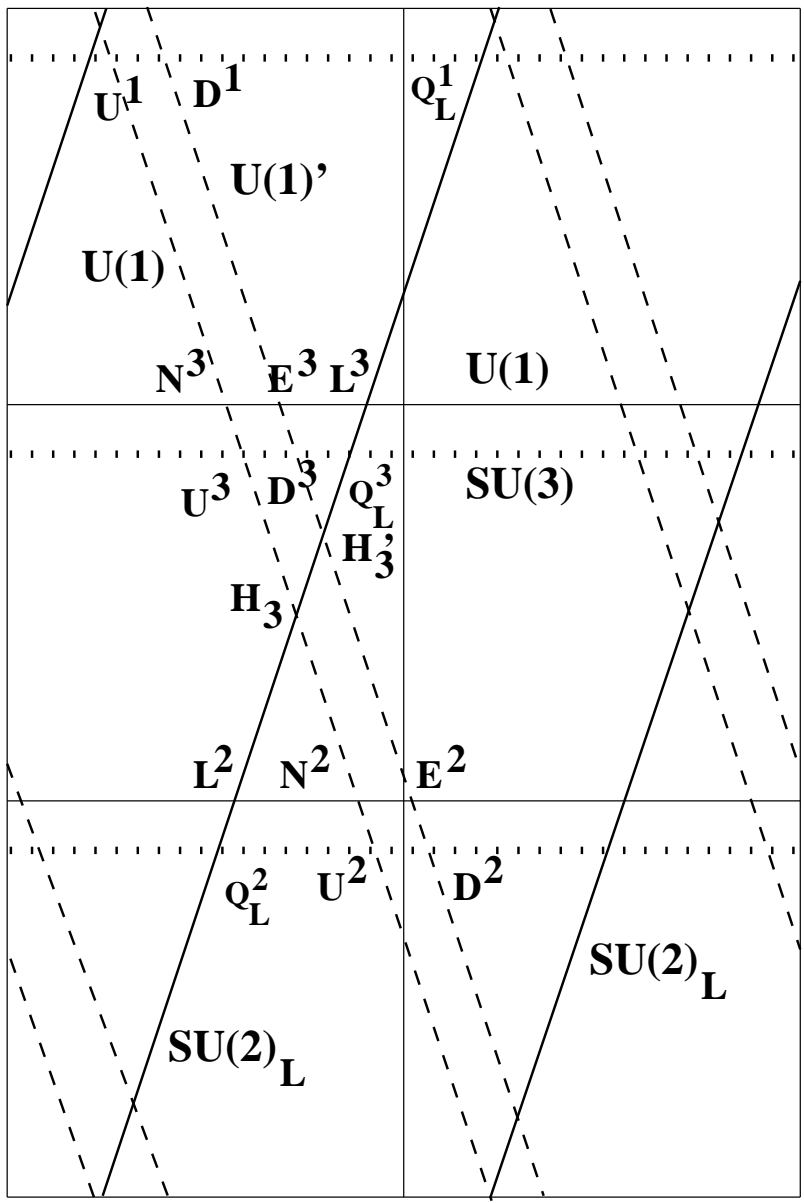}
\end{center}
\caption[]{\small
A standard model-like  configuration obtained from that in Fig.~\ref{inter17} by
splitting the $SU(2)_R$ D4-branes into two parallel $U(1)$-branes.
Now the size of the triangles corresponding to u- and d-quark Yukawa
couplings are different.}
\label{phen4}
\end{figure}

In the left-right symmetric models the Yukawa couplings of u-type and
d-type quarks are equal, although the masses are different if the vevs of
the Higgs fields coupling to u- and  d-quarks are different. In the case
of SM  configurations the Yukawa couplings of u- and d-quarks are in
general different. For example, one may consider a SM-like
configuration  obtained from the
left-right model depicted in Fig.~\ref{inter17} by replacing the two
$SU(2)_R$ D4-branes by two parallel branes next to each other, as shown in
Fig.~\ref{phen4}. In this case, the areas of the different triangles
corresponding to u- and d-quark Yukawa couplings are different, leading to
different hierarchical patterns. This example illustrates how the
location of the different branes allows for different patterns (textures)
for fermion masses. It would be very interesting to study the different
general classes of quark and lepton  textures which can be
accommodated in schemes of this type. Notice that the origin of
hierarchies in this class of models is somewhat similar to that suggested
for heterotic orbifolds in ref.\cite{yukorbi} (see also \cite{yukorbi2} ).
For a recent proposal in the context of brane worlds see \cite{ahs}.

\section{Mass scales and nucleon stability}

The models we are considering are in general non-supersymmetric and hence,
we must set the string scale close to the weak scale to avoid a hierarchy
problem. The four-dimensional Planck scale $M_p$ is related to the string
scale $M_s$ and the compact volumes by (see e.g. \cite{imr})
\beq
M_p\ =\ { {2\sqrt{V_2 V_4} }  \over { \lambda _{II} {\alpha '}^2} }
\label{escalas}
\eeq
where $V_4$ is the volume of the compact variety ${\bf B_4}$ transverse to
the torus where the D4-branes wrap and $V_2=R_1R_2|\sin\theta |$ is the
area of the torus. In order to have not too small gauge and Yukawa couplings
$\sqrt{V_2}/(\lambda_{II} \alpha')$ cannot be very large. Still, one can obtain
the required value for $M_p$ by appropriately choosing a large  value for
$V_4$. In particular, setting the string scale $M_s=1{\rm -}10$ TeV , one
should choose $V_4\approx  10^{16}{\rm -} 10^{10} \; ({\rm GeV})^{-4}$.
For isotropic compactifications, this requires $M_c\approx 3\times
10^{-4}-10^{-2}$ GeV, but this is not the only choice. In fact, two of the
dimensions inside ${\bf B}_4$ could be kept of order the string length,
while the remaining two are taken in the millimeter range, leading to a
phenomenology similar to some  brane-world scenarios considered in the
recent literature.

One of the main problems  for the construction of brane-worlds with a low
scale of order 1-10 TeV is proton stability. If the fundamental scale of
the theory is that low, one expects (unless some symmetry forbids it) the
existence of four-fermion dimension six operators mediating proton decay,
which would be suppressed only by powers of $1/M_s^2$. Interestingly
enough, nucleon decay is automatically forbidden (to all orders in
perturbation theory) in intersecting brane world models. In order for
proton decay to proceed, there must be an effective operator involving
three incoming quarks and no (net) outgoing ones. In our case,
this would require a string amplitude, with e.g. the topology of a disk,
with boundary on the intersecting D-branes, and involving just three
vertex operator insertions associated to the quarks. These arise at
intersections of the $SU(3)$ branes with some other $SU(2)$ or $U(1)$
stack of branes. On the world-sheet boundary, each such insertion changes
the worldsheet boundary conditions from those associated to $SU(3)$ branes
to those associated to $SU(2)$ or $U(1)$ branes (or viceversa). Hence, any
amplitude must involve an even number of such insertions, so there is no
disk configuration which can contribute to proton decay. The argument in
fact is valid for other string worldsheet topologies, with any number of
holes and boundaries, hence the result is exact to all orders in
perturbation theory.

In other words, the above argument applied to any stack of branes shows
that there is an exact discrete symmetry $({\IZ_2})^K$, where $K$ is the
number of brane stacks. Under this symmetry, any state arising from an
open string stretched between the $i^{th}$ and $j^{th}$ stacks of branes
is odd under the $i^{th}$ and $j^{th}$ $\IZ_2$'s, and even under the rest.
The $\IZ_2$ associated to the $SU(3)$ stack of branes prevents proton
decay. Notice that Higgs scalars are neutral under this $\IZ_2$, hence
their vevs do not break this symmetry. These discrete symmetries are
expected to be broken by non-perturbative effects, but their violations
are presumably negligible.

Thus the nucleon is stable in this kind of brane intersection models. This
is a remarkable fact, which is important for scenarios in which the string
scale is close to the weak scale, say at $M_s\propto 1-10$ TeV. Let us
also emphasize that this automatic proton stability is not generic in
other brane world scenarios, such as D3-branes at singularities \cite{aiqu},
but depend on the particular model considered. This feature makes the
intersecting brane world scenario a very interesting proposal.

\section{Low energy spectrum and signatures at accelerators}

The models we are considering have standard quarks and leptons, arising at
the intersections, but are non-supersymmetric and in general squarks and
sleptons are not present. However, the models typically contain extra
particles beyond the content of the minimal SM, which can be rather light.
In this Section we review the main type of extra particles  present in
generic models of this type.

\smallskip

1) {\bf  Excited KK gauge bosons}

The gauge interactions of the standard model are sensitive to the presence
of the toroidal extra dimensions around which the D-branes wrap. Hence in
these models {\it there are Kaluza-Klein replicas of gluons and electroweak
gauge bosons}. In our models of D4-branes, these Kaluza-Klein gauge-boson
excitations have masses (for a general torus metric) given by :
\beq
M^i_{KK} \ =\ {|k|\over {\sqrt{n_i^2R_1^2
+ 2n_im_i \cos\theta R_1R_2 +  m_i^2R_2^2}}}\ \;\;\; {\rm with}\; k\in\IZ
\label{kk}
\eeq
where $i$ labels the different stacks of branes. This formula is interesting
because it can be used to relate the masses of the Kaluza-Klein replicas to
the gauge coupling constants in (\ref{coup})
{\it at the string scale} . Indeed, masses of KK states
are integer multiples of
\beq
 M^i_{KK} \ =\    { {2\alpha_i(M_s)}\over {\lambda_{II} } } M_s
\label{kk2}
\eeq
Thus these replicas are expected to be lighter than the string scale
for  $(\lambda _{II}/2 )\geq \alpha _i$. The expression (\ref{kk2}) also
shows that the  masses of the KK replicas are on the ratios of the
fine structure constants (at the string scale) for the corresponding gauge
bosons. Thus the electroweak excited W's , $\gamma $ and Z's will be in
general the lightest KK modes, and could be the first experimental
signature of extra dimensions (see e.g. \cite{aab}).

Notice that if the excited gauge bosons are relatively light, one has to include
their effect in the running of the gauge coupling constants from $M_s$
down to the electroweak scale. The effect of these excited gauge bosons
would be to make the $SU(3)$ and $SU(2)$ inverse couplings
 to decrease faster as we increase the energies. The overall effect
of this particular contribution
would be analogous to the accelerated running suggested in \cite{ddg}.

\smallskip

2) {\bf Excited gauge bosons from windings }

Depending on the values of the radii $R_1,R_2$ and the wrapping numbers
$(n_i,m_i)$, some string winding states may be below the string scale.
Indeed, for the case of branes {\it multiply wrapped } around  $R_{1,2}$,
there may be open strings stretching between different pieces of the brane
in the fundamental region. For example, there exist such states associated
to open strings stretched between the $SU(2)_R$ D4-brane lines in
Fig.~\ref{phen1}, or between the $SU(2)_L$ or $U(1)'$ D4-brane lines in
Fig.~\ref{phen3}. These states are massive excited gauge bosons in the
corresponding brane, with masses proportional to the separation of the
different pieces of the D4-brane under consideration. The masses of these
winding modes are (for coprime $n, m$)
\beq
M_{\rm stretch}^i\ =\  2\pi p M_s^2 {{R_1R_2|\sin\theta | }\over
{\sqrt{n_i^2R_1^2 +2n_im_i\cos\theta R_1R_2  +  m_i^2 R_2^2}  }  }
\label{stretch}
\eeq
with $p$ a positive integer. Thus, for large wrapping numbers $n_i,m_i$ or
small radii $R_{1,2}$ or  $sin\theta $ some modes may be below the string
scale. Notice that, unlike the KK modes, these states are stringy in
nature, and hence their mass {\it depends explicitly on the string scale}.
For relatively small radii (and for the case of multi-wrapped D4-branes)
these excited gauge bosons may be lighter than the corresponding KK mode
(see also \cite{winding}),
so that either one or the other may be  lighter than the string scale. In
particular, for the case of a square torus ($R_1=R_2$, $\cos\theta =0$)
one can derive the bound for the KK and winding replicas of each gauge
boson,
\beq
M_{\rm stretch}^iM^i_{KK}\ = \ 2\pi (n_i^2+m_i^2)^{-1}\ M_s^2
\label{noloose}
\eeq
so that one or the other could be found at accelerators before reaching
the string threshold.

Unlike the gauge sector, quarks and leptons are localized in the six extra
dimensions and do not have this type of KK excitations. Consequently,
their interactions do not conserve KK quantum numbers, i.e. there exist in
principle couplings of the type $q{\bar q}\rightarrow G^*,W^*,B^*$, of
quarks to KK excitations of gauge bosons (see \cite{aab} and references
therein). Thus KK excitations need not be produced in pairs. Similar
statements can be made about the winding states.

\smallskip

3) {\bf Gonions: Light string excitations of chiral fields}

We have described how the groundstates of open strings stretched between
intersecting branes give rise to chiral fermions. There are also additional
(vector-like) states corresponding to excited open strings (with oscillator
excitations) stretched between the intersecting branes \cite{bdl}. Such modes
are also localized at the vicinity of the intersection. They give rise to
towers of excited states, with spacing controlled by the intersection angle
times the string scale (rather than any compactification scale),
and  are somewhat new in their behaviour. To distinguish them from
the KK and winding excitations of the gauge bosons, we call these
fields {\it gonions}, being associated to branes at angles. There may exist
gonions with spin=1/2,0 and 1. At {\it all} the intersections there are
in general fermionic (vector-like) gonions with masses given by
\beq
m^2_{ij}({\rm fermion})\ =\  q { { | \alpha _{ij}| }\over {\pi}  } M_s^2
\label{gonionf}
\eeq
where $q>0$ is an integer and $\alpha_{ij}$ is the angle formed between the
corresponding pair of branes.
On the other hand at {\it some} of the intersections (concretely, at
those at which Higgs-like fields reside, see sections 7,8) there are in
addition scalar and vector gonions with masses
\beq
m^2_{ij}({\rm scalar})\ =\ (q-1/2){{| \alpha _{ij}| }\over {\pi}} M_s^2
\;\; ;\;\;
m^2_{ij}({\rm vector})\ =\ (q+1/2) {{ | \alpha _{ij}| }\over {\pi}  }
M_s^2 \
\label{gonions}
\eeq
where $q$  is a non-negative integer \footnote{For $q=0$ there are
tachyons which will be discussed in Section 8. They are associated to
Higgs-like fields.}. Thus, the size of these masses depends on the
intersection angles. We will argue in section 8 that these angles may be
relatively small, in order to suppress the weak scale relative to the string
scale. Notice that, the intersection angle $\alpha_{ij}$ depends on the
shape of the torus,
\beqa
\cos\alpha_{ij}\ & = & \ {{g_{ab}v_i^av_j^b }\over {|v_i||v_j|}}\ =
\nonumber \\
& = &  \ {{a^2n_in_j +a \cos\theta (n_im_j+n_jm_i) +m_im_j} \over
{\sqrt{ (an_i)^2 + 2an_im_i\cos\theta +m_i^2} \;
\sqrt{ (an_j)^2 + 2an_jm_j\cos\theta +m_j^2} }  }
\label{angulillo}
\eeqa
where $v_i=(n_i,m_i)$ and $a=R_1/R_2$. Thus, e.g. for $\theta$ close to
$\pi$, the angle $\alpha _{ij}$ becomes close to zero. So, if
$\alpha_{ij} M_s^2 $ is of order the weak scale, one should see the first
excited (vector-like) replicas of the observed quarks and leptons not much
above the weak scale. These masses will be generation independent, but
differ from one type of standard model fermion to the other since
their masses are proportional to the corresponding intersection angles.

These gonion  excitations of the chiral fields in the intersections are
the most likely signature of the present scheme at accelerators. They have
the same quantum numbers under the gauge group as the corresponding quark or
lepton living at the corresponding intersection. Thus, for example,
coloured gonions should be produced by gluon fusion at a hadronic collider,
and would look very much like new vector-like quark generations with
generation independent masses. In addition all type of gonions have couplings
to the ordinary quarks and leptons which will be of order of the usual Yukawa
couplings. For example, a scalar or vector gonion in the same intersection
as a Higgs field, will have couplings to quarks and leptons proportional
to the corresponding Yukawa couplings. This is because the coupling would
be proportional to $\exp(-A_{ijk})$, with $A_{~ijk}$ the area of the
worldsheet stretched among the gonion and the two fermion intersections, very
much like in standard Yukawa couplings. Thus, bosonic gonions will typically
decay into third generation quarks and leptons. Again, note that if these
gonions have masses not much above the weak scale (as suggested in section
8), they will contribute to the running of the gauge couplings in between
the weak and the string scales.

\smallskip

4) {\bf Extra massless states in the brane bulk}

The massless sector of each of the D4-branes of course includes the gauge
bosons of the corresponding gauge group, but may contain extra particles.
In particular, although the complete theory is non-supersymmetric due to
the presence of the intersections, the gauge sector living on the bulk of
the D4-branes (i.e. within the brane, but away from the intersections) may
be supersymmetric, even with $\NN=2$ or $\NN=4$ supersymmetry. In this
case, besides the gauge bosons, there exist fermionic and/or bosonic
partners transforming in the adjoint of each gauge group. The presence or
not of these enhanced SUSY sectors depends on the geometry of the
transverse compact variety ${\bf B_4}$ \footnote{In Section 7 we construct
specific string models in which $B_4=\IT^4/\IZ_N$, with an enhanced
$\NN=2$ supersymmetry in the bulk of the D4-branes. Analogous models with
$\NN=0$ may be obtained by performing a $\IZ_N$ twist breaking all SUSY's.
See \cite{afiru} for details.}.

The simplest possibility from the phenomenological perspective is having
no SUSY in the bulk. Even in this case, there may be additional scalars
and vector-like fermions transforming in the adjoint of each gauge group,
and massless at tree level. Indeed, the presence of these scalars would
signal the possibility of separating the branes within a stack (i.e. like
the two $SU(2)_R$ D4-branes of left-right symmetric models) into a set of
parallel branes. They would lead to e.g. $SU(3)$ octet scalars and
$SU(2)_L$ triplet scalars. Although massless at the tree-level, both
scalars and fermions would acquire one-loop masses, see eq.
(\ref{massloop}), of order $\approx \alpha_i M_s$. If present, they could
also provide interesting signatures at colliders.

In addition to the above signatures, one may have the standard signature of
extra dimensions  of graviton emission to the bulk (corresponding to
the large transverse space ${\bf B_4}$), which has been extensively
analyzed in the literature \cite{graviton}. Obviously, if the string scale
is reached, explicit string modes would be accessible. However, as pointed
out above, in the present scenario the KK/winding excitations of gauge
bosons, and gonion excitations of chiral fields are expected to be
lighter, and much more accessible. A detailed phenomenological analysis of
their production at colliders would be interesting.

\section{Explicit string models}

In this section \footnote{Readers not familiar with technicalities of
string theory may skip to the following section.} we would like to present
specific Type IIA string models, with D4-branes wrapping on a torus,
yielding structures very similar to the ones sketched in the previous
sections. 

The kind of configurations we consider here have been recently studied
in \cite{afiru}, to which we refer the reader interested in the more
technical details. Here we will merely present several of these string
constructions, providing explicit realizations of the scenario discussed
in section 2. As explained in \cite{afiru}, D4-branes in flat
space lead to non-chiral matter content in their intersection. One is
therefore led to consider D4-branes (with one direction wrapped on
one-cycles in a two-torus) sitting at singular points in a transverse
space, which we take to be ${\bf B_4}= (\IT^2)^2/\IZ_N$.

For concreteness we center on $\IZ_3$ orbifolds (extension to the general
case being straightforward \cite{afiru}), generated by a geometric action
$\theta$ with twist vector $v=\frac{1}3(1,-1,0,0)$. We consider $K$ different
stacks of D4-branes, each one containing $N_i$ branes, with wrapping
numbers around the 2-torus given by $(n_i,m_i)$. We set the four
transverse coordinates of the D4-branes at the fixed point at the origin
in $(\IT^2)^2/\IZ_3$. The $\IZ_3$ action may be embedded in the $U(N_i)$
gauge degrees of freedom of the $i^{th}$ stack of D4-branes, through a
unitary matrix of the form
\beq
\gamma_{\theta,i} \ = \ \diag(\id_{N_i^0},e^{2\pi i\frac 13} \id_{N_i^1},
e^{2\pi i\frac 23} \id_{N_i^2})
\label{chanpaton}
\eeq
with $\sum_a N_i^a=N_i$. Due to this twist the initial gauge group
$\; \prod_{i=1}^K U(N_i) \;$ is broken to $\quad \prod_{i=1}^K
\prod_{a=1}^3 U(N_i^a)$.

Cancellation of twisted tadpoles in the theory imposes the constraints
\footnote{We do not impose cancellation of untwisted tadpoles,
assuming they are properly cancelled by an additional set of D4-branes
away from the origin in $(\IT^2)^2/\IZ_3$. Such extra branes do not change
the field theory spectrum in the sector at the origin, and hence are
irrelevant for our discussion.}
\beqa
\sum_{i=1}^K n_i \, \Tr \gamma_{\theta^k,4_i} = 0 \quad ; \quad
\sum_{i=1}^K m_i \, \Tr \gamma_{\theta^k,4_i} = 0
\label{tadpofour}
\eeqa
These conditions guarantee, as usual, the cancellation of gauge anomalies.
At the intersections of the different D4-branes, there appear massless
fermions transforming under $\prod_{i=1}^K \prod_{a=1}^3 U(N_i^a)$ as
\cite{afiru}
\beq
\quad \sum_{i<j} \sum_{a=1}^3 \;  I_{ij}\times
[\; (N_i^a,{\ov  N}_j^{a+1}) + (N_i^a,{\ov  N}_j^{a-1})
-2(N_i^a,{\ov  N}_j^a)
\; ]
\label{espec}
\eeq
with the usual convention for negative multiplicities (see footnote 5).
One easily checks that tadpole cancellation conditions indeed imply that
this fermion spectrum is free of non-Abelian gauge anomalies. Concerning
mixed $U(1)$ anomalies, some of the $U(1)$ gauge symmetries have triangle
anomalies, as is often the case in string theory constructions.
The theories are nevertheless consistent, due to the cancellation of the
anomaly by a generalized Green-Schwarz mechanism, involving twisted closed
string states. The corresponding gauge bosons become massive, with mass of
the order of the string scale, by  combining with certain twisted closed
string scalars, whereas the orthogonal linear combinations are anomaly-free
and remain massless (see \cite{afiru} for details). Armed with the above
information, we can now construct explicit string compactifications
similar to the examples given in section 2.

Before showing specific models, notice that, once the wrapping numbers
have been specified, an infinite number of models can be constructed  by
acting on all the wrapping vectors $(n_i,m_i)$ with (the same) $SL(2,\IZ)$
transformation \footnote{Matrices of the form,
\beq
C \ =\
\left( \begin{array}{cc}
a & b \\
c & d
\end{array}
\right)
\eeq
where $a,b,c,d$ are integers and $det(C) = 1$}. This kind of transformations
preserves the intersection numbers between different sets of branes, i.e. the
chiral spectrum. Distances are also preserved if the metric transforms
accordingly \footnote{If $g_A$ is metric of the original torus, the
transformed metric should be of the form $g_B = (C^{-1})^T g_A C^{-1}$.}.
Two models in the same $SL(2,\IZ)$ family represent the same physics: the
spectrum is related to the intersection matrix and the masses are related
to the metric of the torus.

It is therefore interesting to classify all non-equivalent models leading,
to the same intersection matrix. This number turns out to be just the sum
of the divisors of the number of generations, e.g. for three generations
there are four non-equivalent families of three generation models ($1+3$).
To obtain all non-equivalent families with a given intersection matrix,
one would proceed as follows.
\begin{itemize}

\item Consider a pair of D-brane stacks, $i$ and $j$, with intersection
$I_{i,j}$, and find all non-equivalent pairs of wrapping numbers with such
intersection.

\item For each fixed choice of wrapping numbers, the remaining wrapping  
numbers are determined by imposing the intersection numbers with $i$ and
$j$, which are now linear equations.

\item Finally, one should check the intersections among branes different 
from $i$ and $j$. Also, solutions with non-integer wrappings should be
rejected.

\end{itemize}

With an intersection number $I_{ij} > 0$, we can use $SL(2,\IZ)$ to bring
the wrappings of the stacks $i$ and $j$ to the form $(n_i,0)$ and
$(n_j,m_j)$, with $n_i > 0$, $m_j > n_j \geq 0$, and $n_i m_j=I_{ij}$.
The number of solutions is just the sum of all the divisors of $I_{ij}$.
Each solution then determines the remaining wrapping numbers in terms of
intersection numbers. Leaving a full study of the characteristics of the
different families, we turn to studying a couple of examples of three
generation models.

\medskip

{\bf Example 1}

Consider five sets of D4-branes with multiplicities $N_1=3$, $N_2=2$,
$N_3=2$ and $N_4=N_5=1$, and wrapping numbers
\beq
N_1:(1,0) \;\; ;\;\; N_2:(1,3) \;\; ; \;\; N_3:(1,-3) \;\; ;\;\; N_4:(-1,0)
\;\; ;\;\; N_5=(3,0)\ .
\label{lrs1}
\eeq
Notice that this choice is identical to the one in example 1 of Section 2,
except for one additional D4-brane. The latter will be required in the
present example in order to cancel the twisted tadpole conditions, and
render the string configuration consistent. The twists acting on CP
indices are taken to be
\beqa
\gamma_{\theta,1} \ &=& \ \id_{3} \nonumber \\
\gamma_{\theta,2}=\gamma_{\theta,3} \ &=& \alpha \id_2
\nonumber \\
\gamma_{\theta,4}\ &=& \ \alpha  \nonumber \\
\gamma_{\theta,5}\ &=& \ \alpha ^2
\label{gamos}
\eeqa
where $\alpha=\exp(2\pi i/3)$. One can easily check that these choices of
wrapping numbers and CP twist matrices verify the tadpole cancellation
conditions (\ref{tadpofour}). The gauge group is $U(3)\times U(2)_L\times
U(2)_R\times U(1)_4\times U(1)_5$. Using (\ref{inter}) and (\ref{espec}),
one easily obtains the massless chiral fermion spectrum displayed in
Table~\ref{tabpslr}.

\medskip

\begin{table}[htb] \footnotesize
\renewcommand{\arraystretch}{1.25}
\begin{center}
\begin{tabular}{|c||c|c|c|c|c|c||c|c|}
\hline Intersection &
 Matter fields  &  $Q_1$  & $Q_2 $ & $Q_3 $ & $Q_4$ &
   $Q_5$
& $B-L$  & X  \\
\hline\hline (12) & $3(3,2,1)$ & 1  & -1 & 0 & 0 & 0 & 1/3  & 0 \\
\hline (13) &   $3(\bar 3,1,2)$ & -1  & 0  & 1 & 0 & 0 & -1/3 & 0 \\
\hline (23) &  $12(1,2,2)$ & 0  & 1  &  -1 & 0 & 0 & 0  & 0 \\
\hline (24) &  $6(1,2,1)$ & 0 & -1 & 0 &  1  & 0 &   1  & 1 \\
\hline (34) &  $6(1,1,2)$ & 0 & 0  & 1 &  -1  & 0 &  -1  &  -1 \\
\hline (25) &  $9(1,2,1)$ & 0 & -1 & 0 &  0  & 1 &  -1  & -2/3 \\
\hline (35) &  $9(1,1,2)$ & 0 & 0 & 1 &  0  & -1 &  1 &  2/3  \\
\hline \end{tabular}
\end{center}
\caption{\small Spectrum of the $SU(3)\times SU(2)_L\times SU(2)_R$
model. We present the quantum numbers of the chiral fermions  under the
$U(1)^5$ group, as well as the charge under the $B-L$ linear combination
and the  anomaly-free generator $Q_X$.
\label{tabpslr} }
\end{table}

\medskip

Non-abelian cubic anomalies automatically cancel, while there are two
anomalous $U(1)$'s which become massive. The
diagonal sum of the five $U(1)$ generators is anomaly-free, but
actually it decouples since all
particles have zero charge under it. 
In addition there are  two anomaly-free linear
combinations 
\beqa
Q_{B-L}\ &=& \ -{2\over 3}Q_1-Q_2-Q_3-2Q_5^D   \nonumber \\
Q_X \  & = & \ Q_4  - {2\over 3} Q_5^D
\label{bmenosl}
\eeqa
We have displayed the charge under these two generators in
Table~\ref{tabpslr}.
 The first linear combination plays the role of $B$-$L$
symmetry. The model contains three quark-lepton chiral generations, plus
some additional vector-like leptons \footnote{Notice that the number of 
generations arises from the intersection number between the cycles
(\ref{lrs1}), and is completely unrelated to the order of the
orbifold group $\IZ_3$.}.
There is an additional subtlety here concerning the fifth
brane  with wrapping $(3,0)$. It turns out that 
whenever $n$ and $m$ are not coprime as in this  case, 
a brane gives rise not to a single $U(1)$ field but to several
copies. In the present case a brane wrapping with 
$(n,m)=(3,0)$ gives rise to three $U(1)$ fields
with generators $Q_5^a$, $a=1,2,3$  corresponding to
open strings stretching between the first wrapping of the brane
and the first,second and third wrappings. Thus in addition 
to the above two anomaly-free $U(1)$'s 
(with $Q_5^D=\sum_a Q_5^a$) there are other two ($Q_5^1-Q_5^2$ and
$Q_5^2-Q_5^3$) 
which only 
couple  to the fields in the intersections $(25)$ and $(35)$
which we have not desplayed in the table.

Comparing with example 1 in Section
2, besides these extra leptons there are the additional
anomaly-free $U(1)$'s mentioned above. They  arise from the
additional D4-brane we have
introduced for technical reasons, namely in order to achieve cancellation
of twisted tadpoles. Their presence should not be considered as a necesary
consequence of the present scenario but rather from 
its particular realization.

\medskip

{\bf Example 2}

Consider five different stacks of D4$_i$-branes, with multiplicities
$N_1$=3, $N_2=2$ and $N_3=N_4=N_5=1$, and wrapping numbers \footnote{Indeed,
there are other three $SL(2,\IZ)$ families with the same intersection
numbers,
\beqa
N_1:(1,0) \;\; ;\;\; N_2:(0,3) \;\; ; \;\; N_3:(1,-3) \;\; ;\;\;
N_4:(2,-3) \;\; ;\;\; N_5=(3,0)\  \nonumber \\
N_1:(1,0) \;\; ;\;\; N_2:(2,3) \;\; ; \;\; N_3:(-1,-3) \;\; ;\;\;
N_4:(0,-3) \;\; ;\;\; N_5=(3,0)\  \nonumber \\
N_1:(3,0) \;\; ;\;\; N_2:(0,1) \;\; ; \;\; N_3:(3,-1) \;\; ;\;\;
N_4:(6,-1) \;\; ;\;\; N_5=(9,0)\ .
\eeqa
Notice however that only the second family leads to a gauge group
$SU(3)\times SU(2)\times U(1)^n$. In the remaining families, color or weak
branes are multiply wrapped and lead to a replication of the corresponding
gauge factor.}

\beq
N_1:(1,0) \;\; ;\;\; N_2:(1,3) \;\; ; \;\; N_3:(0,-3) \;\; ;\;\;
N_4:(1,-3) \;\; ;\;\; N_5=(3,0)\ .
\label{sms1}
\eeq
This choice is similar to example 2 in Section~2, differing only in the
introduction of one additional D4-brane, required to achieve cancellation
of twisted tadpoles in the model. The twists acting on CP indices are
taken to be
\beqa
\gamma_{\theta,1} \ &=& \ \id_{3} \nonumber \\
\gamma_{\theta,2}&=& \ \alpha \id_2
\nonumber \\
\gamma_{\theta,3}= \gamma_{\theta,4}\ &=& \ \alpha  \nonumber \\
\gamma_{\theta,5}\ &=& \ \alpha ^2
\label{gamis}
\eeqa
Again, one can easily check that these choices of wrapping numbers and CP
twist matrices verify the tadpole cancellation conditions
(\ref{tadpofour}). The gauge group is  $U(3)\times U(2)_L\times
U(1)_3\times U(1)_4\times U(1)_5$. From (\ref{inter}) and (\ref{espec}),
the spectrum of chiral fermions is easily computed, and the result is
shown in Table~\ref{tabpssm}.

There are two anomaly-free $U(1)$ linear combination (apart from
the diagonal one, which decouples) given by
\beqa
Q_{Y}\ & =&\ -{1\over 3}Q_1 -{1\over 2}Q_2 -Q_4-Q_5  \nonumber \\
Q_{X}\ & = & Q_3 -Q_4-{2\over 3} Q_5
\label{bhyper}
\eeqa
Table~\ref{tabpssm} also provides the charges under these linear
combinations. Interestingly, we see that the first of these generators can
be identified with standard weak hypercharge. Again, the model contains
three quark-lepton generations plus some vector-like leptons and
additional $U(1)$ gauge factors. As happened in the left-right symmetric
model, in the present case the third brane
 with wrapping $(0,-3)$ and the fifth with $(3,0)$ 
give rise to 2+2 additional anomaly-free $U(1)$'s whose  
charge we have not desplayed in the table.
Comparing with example 2 of Section 2, we
find the model is very similar , the differences being due to the
presence of an additional D4-brane, which we have been forced to
introduce in order to satisfy twisted tadpole cancellation conditions.

\begin{table}[htb] \footnotesize
\renewcommand{\arraystretch}{1.25}
\begin{center}
\begin{tabular}{|c||c|c|c|c|c|c||c|c|}
\hline Intersection &
 Matter fields  &  $Q_1$  & $Q_2 $ & $Q_3 $ & $Q_4$ &
   $Q_5$
&  Y  & X \\
\hline\hline (12) & $3(3,2)$ & 1  & -1 & 0 & 0 & 0 & 1/6 & 0  \\
\hline (13) &   $3(\bar 3,1)$ & -1  & 0  & 1 & 0 & 0 & 1/3 &  1  \\
\hline (14) &   $3(\bar 3,1)$ & -1  & 0  & 0 & 1 & 0 & -2/3 &  -1 \\
\hline (23) &  $6(1,2)$ & 0 & 1 & -1 &  0  & 0 &   -1/2  & -1  \\
\hline (24) &  $12(1,2)$ & 0 & 1 & 0 &  -1  & 0 &  1/2   &  1  \\
\hline (25) &  $9(1,2)$ & 0 & -1 & 0 &  0  & 1 &  -1/2 & -2/3  \\
\hline (34) &  $6(1,1)$ & 0 & 0 & -1 &  1  & 0 &  -1  &  -2  \\
\hline (35) &  $9(1,1)$ & 0 & 0  & 1 & 0   & -1 &  1  & 5/3 \\
\hline (45) &  $9(1,1)$ & 0 & 0  & 0 & 1   & -1 &   0 & -1/3    \\
\hline \end{tabular}
\end{center}
\caption{\small Spectrum of a standard model. We present the quantum
numbers of the chiral fermions  under the $U(1)^5$ group, as well as the
hypercharge linear combination and the additional $Q_X$ generator.
\label{tabpssm} }
\end{table}

The above two examples illustrate how the general properties described in
the previous sections may in fact be obtained in the context of string
theory. Although in these particular examples, due to technical reasons,
 we were forced to add an extra brane, which led to
 extra $U(1)$'s and additional leptons, our discussion of gauge and Yukawa
couplings, structure of mass scales, proton stability, and the possible
presence of light KK/winding  gauge boson excitations 
and gonions remains valid for these
explicit string examples. 

As an illustration we can estimate the possible values of coupling constants
as discussed in section 3. Recall that, since the hypercharge generator
(\ref{bhyper}) involves the additional D4-brane, not present in
(\ref{bhypertoy}) in the toy model in Section~2, we must replace
$|(n_4,m_4)|\rightarrow |(n_4,m_4)|+|(n_5,m_5)|$ in (\ref{coupsm}). For
instance, by choosing $\cos{\theta} \simeq -1$ and $R_2/R_1=1.57$, we obtain
the $\alpha_i$'s are in the ratios $1: 0.27 : 0.09$ which coincide, within
less than 6 \% with experimental ratios  $1: 0.268 : 0.0861$. A more
precise determination of low-energy would require taking into account the
effect of different thresholds as discussed above. In any event, as
claimed in Section~3, there seems to be enough freedom to reproduce
experimental values of coupling constants in the present setup.

\medskip

The specific examples discussed in this section have however a potential
problem, regarding the scalar sector, as pointed out in \cite{afiru}. In the
class of models with D4-branes  wrapping  on $\IT^2\times (\IT^2)^2/\IZ_N$
that we are discussing, there are tachyonic scalars appearing at {\it
some} of the D4-brane intersections. In particular, for a general
set of D4-branes at a $\IZ_3$ orbifold, there appear complex scalars at
intersections involving D4-branes with the same eigenvalue in the CP twist
matrix $\gamma_{\theta^k,4_i}$. They transform under $\prod_{i=1}^K
\prod_{a=1}^3 U(N_i^a)$ as
\beq
\quad \sum_{i<j} \sum_{a=1}^3 \;  I_{ij}\times
\; (N_i^a,{\ov  N}_j^{a}) \;
\label{tac}
\eeq
Their masses are given by
\beq
M_{ij}^2\ =\ - {{M_s^2}\over 2} |{ {\alpha _{ij}}\over {\pi} } |
\label{mtac}
\eeq
where $|\alpha _{ij}|$ is the angle at which the corresponding pair of
D4-branes intersect on the torus. Thus the model contains tachyons at
those intersections. Their properties are discussed in more detail in
next Section.

\section{Tachyons and electroweak symmetry breaking}

In the specific string compactifications described in previous section,
besides the chiral fermions present at every intersection, there exist
complex scalars at {\it some} of them. For example, as one can read from
(\ref{tac}), in the standard model example 2 of previous Section there are
complex scalars in the intersections (23), (24) and (34), transforming as
$(1,2)_{-1/2}$ , $(1,2)_{1/2}$ and $(1,1)_{-1}$ under $SU(3)\times
SU(2)\times U(1)_Y$ respectively. In the case of the left-right model,
example 1, there are complex scalars at the same intersections,
transforming as $(1,2,2)$, $(1,2,1)$ and $(1,1,2)$ under $SU(3)_c\times
SU(2)_L\times SU(2)_R$.

As we mentioned in the previous chapter their masses are given by
(\ref{mtac}), and hence they are tachyonic. This signals an instability of
the brane configuration which tends to favour the alignment of the
D4-branes along parallel directions. On the other hand, the fact that in
these examples some of the {\it  tachyons have precisely the quantum
numbers of Higgs fields} suggests that perhaps what these tachyons indicate
is some stringy version of a Higgs mechanism \cite{afiru} (see also
\cite{bachas} for an early proposal of the SM Higgs as tachyon, in a
different (but related) context). Since many of the theoretical aspects of
the tachyon potential and dynamics are still under study (see
\cite{tachsen,sft,sftoy} for some recent references on tachyon
condensation in brane-antibrane systems), our discussion in this Section is
tantalizing, but to some extent qualitative.

A possible obstacle for this interpretation is that naively
tachyonic masses are of the order of the string scale. In the
case of the Standard Model, that would require a string scale of
the order of the weak scale, a possibility not consistent with
experimental observations. The situation would be  better for the
case of tachyonic $SU(2)_R$ doublets in left-right symmetric
models, since $SU(2)_R$ breaking at the TeV scale would require a
string scale in the region 1-10 TeV, which can be achieved without
contradiction with experiment.

However, the situation is better, even for SM configurations. In fact, as
follows from (\ref{mtac}), the mass of tachyons may be substantially
smaller than the string scale if the intersection angles $\alpha _{ij}$
are sufficiently small (but non-vanishing, so that the branes intersect to
yield a chiral model). In particular, by varying the shape (complex
structure) of the torus one can make all these angles arbitrarily small.

In particular consider the case of a squashed torus with $\theta $ close
to $\pi$, so that $\cos\theta = -1+\epsilon ^2/2$. In that case one can
check using (\ref{angulillo}) that the angles between the different
D4-branes are proportional to $\epsilon$, and hence be made arbitrarily
small. In particular it is easy to find in that limit:
\beq
M_{ij}^2\ = \ - {{M_s^2}\over 2}  |{ {\alpha _{ij}}\over {\pi} } |\ =
 \ - {{M_s^2}\over 2} { { a\epsilon \
|I_{ij}|}\over
 {|an_i-m_i||an_j-m_j|} }
\eeq
where, if $m_i\not= 0$,  $a=R_1/R_2>m_i/n_i, m_j/n_j$. Here $I_{ij}$ is
the intersection matrix described in chapter 2. Thus we see that the
size of the negative tachyonic mass may be made arbitrarily low by fixing
$\epsilon$ (or in some cases $a$)  to a sufficiently small value.

In terms of the effective field theory, this negative mass square signals
the breaking of the gauge symmetry. Consider first the SM example 2. The
$SU(2)_L$ doublets at the intersections (24) and (23) and the $SU(2)_L$
singlet at the intersection (34) have masses
\beq
M_{24}^2 \ =\ - {{M_s^2}} { { 3a\epsilon }\over
 {(a^2-9)} } \ ;\
M_{23}^2 \ =\ - {{M_s^2}} { { a\epsilon }\over
 {2(a-3)} } \ ;\
M_{34}^2 \ =\ - {{M_s^2}} { { a\epsilon }\over
 {2(a+3)} }
\label{mhiggs}
\eeq
with $a>3$. Consider for example a value $a=R_1/R_2=10/3$. Then these
negative masses would be in the ratios $54/19 : 3 : 3/19$ respectively.
Thus the negative mass square of the Higgs doublets in the intersections
are much larger than that of the charged singlet and hence standard
electroweak breaking would be preferred \footnote{As discussed below, loop
effects tend to give positive contributions to the scalar masses, which
can easily overcome the tiny tachyonic mass of the singlet scalar.}.

This would certainly be an intriguing origin for electroweak symmetry
breaking. Whereas in the standard model a negative $(mass)^2$ is put by
hand for the Higgs doublet, in the present scheme it appears naturally due
to the presence of tachyons at brane intersections. Hence chirality
and gauge symmetry breaking are linked in these models: chirality requires
intersecting branes, which yield tachyonic modes which in turn trigger
electroweak symmetry breaking.

\begin{figure}
\begin{center}
\centering
\epsfysize=7cm
\leavevmode
\epsfbox{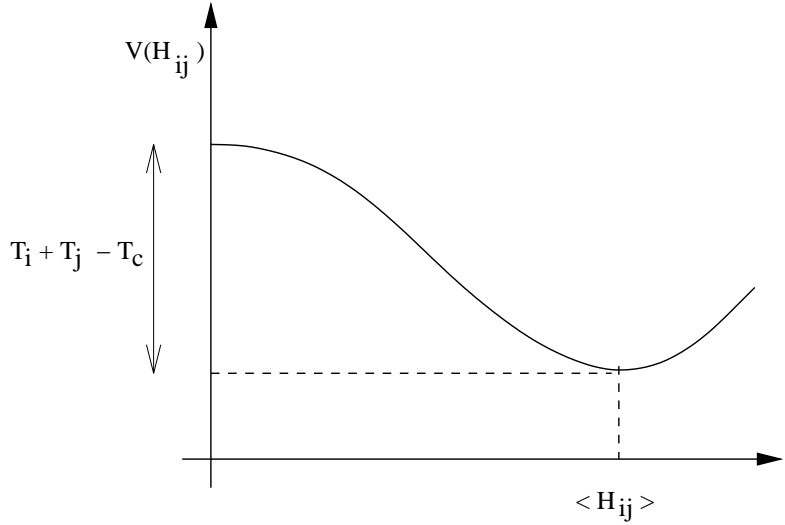}
\end{center}
\caption[]{\small Qualitative form of the tachyon (Higgs) potential
originated by intersecting brane instability.}
\label{tacpot}
\end{figure}

From the point of view of string theory the interpretation goes as follows.
The presence of tachyons in two intersecting D4-branes signal an instability
of the system under recombination of both into a single D4-brane. For
example, consider again the SM construction, example 2 above. There are
two parallel D4-branes with wrapping numbers $(n,m)=(1,3)$ which give rise
to $SU(2)_L$ gauge interactions. They intersect with another brane with
wrapping number $(0,-3)$, and at the intersections we get tachyonic scalars
with masses as in (\ref{mhiggs}). Their presence indicates an instability
of the system against the recombination of e.g. one of the $(1,3)$ branes
with the $(0,-3)$ brane, giving rise to a single D4-brane with wrapping
numbers $(1,3)+(0,-3)=(1,0)$. The string theory construction
shows that the recombination process corresponds to the tachyon field
rolling to a minimum, which is reached in the final configuration. In the
process, the tachyon condensate breaks the gauge symmetry. Namely, the
non-Abelian $SU(2)_L$ generators disappear from the massless spectrum
since there only remains one $(1,3)$ brane instead of two. Thus, with the
{\it tachyon at the minimum of its potential} two intersecting D4-branes
have merged into a single one.

The detailed form of this scalar potential is not known, although the
properties of similar tachyons in brane-antibrane configurations have been
studied e.g. in \cite{tachsen,sft,sftoy}. For instance, adapting the
results in \cite{tachsen}, one concludes that, if a D4-brane $i$ combines
with a D4-brane $j$ to form a combined D4-brane $c$, the depth of the
potential is given by the difference of the D-brane tensions (after
compactification on their corresponding cycles) \footnote{See \cite{gns}
for a similar statement in a different (but related) context.}. That is,
$\Delta V= T_c-(T_i+T_j)$, where \cite{polchi}
\beqa
T_i= \frac{M_S^4}{(2\pi)^4 \lambda_{II}} |(n_i,m_i)| =
M_s^4/(16\pi ^3 \alpha_i(M_s))
\eeqa
and analogously for the branes $j$ and $c$. Here $|(n,m)|$ is the length
(\ref{length}), and $\alpha_i$ the fine structure constant for the
corresponding group. This is schematically shown in Fig.~\ref{tacpot}.

In the regime of small interbrane angles discussed above, the potential
depth is small. Specifically, for the recombination discussed above
$(1,3)+(0,-3)\to (1,0)$, one obtains
\beqa
\Delta V = \frac{M_S^4}{(2\pi)^4 \lambda_{II}} \frac {3 (R_1M_S)}{2(a-3)}
\epsilon^2
\eeqa
so for $R_1$ of order one in string units, $\Delta V$ is of the order of
$\epsilon^2 M_S^4$. Even though the detailed form of the potential is not
known, one can make a rough estimate of the tachyon vev at its minimum (by
computing at which vev the mass term cancels the tension difference) to
be of order $\sqrt{ \epsilon } M_s$.

If we communicate an amount of energy larger than $M_s
\sqrt{\epsilon}$ to the system, the vev of the tachyon becomes irrelevant.
This means that we are able to resolve the combined brane into the
original pair of branes, and produce $W$-bosons. This is certainly a quite
intriguing interpretation of the process of electroweak symmetry breaking
in the standard model \footnote{As pointed out in \cite{afiru} the
tachyon condensation process is analogous to a standard Higgs mechanism
as long as no other gauge symmetry enhancements are available in the
probed energy range. In our case, this would require that other sets of
branes with total wrapping $(1,0)$ are heavier than the considered pair
$(1,3)+(0,-3)$. Suitable choices of geometric moduli lead to this
behaviour.}.

The tachyonic scalar masses given in (\ref{mhiggs})  are
tree-level results. In addition all scalars receive corrections to their
$(mass)^2$ from loop effects. One can estimate those corrections from the
effective field theory. In particular, one gauge boson exchange gives
corrections of order
\beq
\Delta M^2(\mu ) \ =\ \sum_a { {4C_F^a \alpha_a(M_s) }\over {4\pi }} M_s^2
f_a \log(M_s/\mu) \ +\ \Delta M^2_{KK/W}
\label{massloop}
\eeq
where the sum on $a$ runs over the different gauge interactions and $C_F^a$
is the eigenvalue of the quadratic Casimir in the fundamental representation.
Here $\Delta M^2_{KK/W}$ denotes further contributions which may
appear from the KK/W and gonion
 excitations if they are substantially lighter than the
string scale $M_s$. The function $f_a$ is given by
\beq
f_a \ =\  { {2+b_a{{\alpha_a(M_s)}\over {4\pi }} t}\over
{1+b_a{{\alpha_a(M_s)}\over {4\pi }}t } }
\eeq
where $t=2\log(M_s/\mu)$ and $b_a$ are the coefficients of the one-loop
$\beta $-functions. These corrections are positive and may overcome in
some cases the tachyonic masses if the latter are small. Extra KK/winding
excitations may contribute to this effect if they lie between the weak and
the string scales. In particular, notice that since the intersection
angles are small, as suggested above, there will be relatively light {\it
gonion} excited fields, of the type discussed in Section 6, just above the
weak scale, and contributing to one-loop corrections.

In addition, a doublet scalar should have a large Yukawa coupling to the
top quark, giving rise to a negative one-loop contribution to the
$(mass)^2$ of the doublet. This would contribute further to inducing
electroweak symmetry breaking, very much as in the radiative symmetry
breaking mechanism \cite{ir}. A full description of electroweak symmetry
breaking in this class of models would thus require an understanding of
these loop corrections which may compete with the tree-level ones.

\section{Final comments and outlook}

In this paper we have presented a string scenario in which there is one
brane-world per SM gauge interaction. At the intersections of the branes
live the quarks and leptons, which are the zero modes of open strings
close to each intersection. Our original motivation for this proposal was
the fact that brane intersections is one of the few known ways to obtain
chirality in the brane world context in string theory. In addition it
offers an explanation for quark-lepton family replication, since
generically branes can intersect at multiple points.

While studying the proposal we have found a number of interesting aspects
of this scheme. For instance, hierarchical Yukawa couplings naturally
appear due to the fact that the quarks, leptons and Higgs fields are
located at different points in the compact dimensions. The Yukawas are
proportional to $e^{-A_{ijk}}$, where $A_{ijk}$ is the area of the
worldsheet extending among the intersections where the fermions and the
Higgs live. Due to this fact, it is easy to obtain hierarchical results
for the different Yukawa couplings. Next, the models are
non-supersymmetric, but the hierarchy problem may be solved by lowering
the string scale down to $1-10$ TeV, and taking the dimensions transverse
to the branes large enough. Interestingly enough, even though the string
scale is so low, the proton is naturally stable to all orders in
perturbation theory, due to discrete symmetries following from worldsheet
selection rules. The proton is stable because its decay would require an
overall interaction with three incoming quarks and no outgoing ones. Such
process would require worldsheets with an odd number of quark insertions,
which do not exist. Finally, concerning gauge coupling constants, we have
found that they do not unify in this setup, since each brane comes along
with its own coupling constant. However, they may be computed in terms of
the compactification radii, and may be made compatible with the observed
values.

One of the interesting aspects of the intersecting brane-worlds scenario
is that it predicts the existence of certain particle excitations in the
energy region between the weak and the string scales. There are KK (and/or
winding) replications of the gauge bosons, which could be directly produced
at colliders by  quark-antiquark annihilation. In addition there is a new
class of states, which we have baptized as {\it gonions}, which have
masses proportional to the string scale times the intersection angles,
(hence the name gonions). They correspond to excited strings stretched
close to the intersection of two branes. They include massive vector-like
copies of quarks and leptons. In addition there are bosons with spin=0,1
close to some of the intersections. All of them come in towers
starting about the weak scale. It should be interesting to study in more
detail the experimental signatures of these new fields as well as setting
limits on their masses from present data.

Like in many non-supersymmetric models, the spectrum contains scalar
tachyons. Interestingly enough, in the specific string models that we
construct, those tachyons have precisely the quantum numbers of Higgs
fields. Thus it is tempting to propose that these tachyonic states are
just signaling the presence of spontaneous gauge symmetry breaking.
It should be interesting to explore in more detail the theoretical
viability of this exciting possibility.

In this article we have concentrated on the simplest possibility of
D4-branes wrapping at angles on a torus. We would like to emphasize,
however, that most of the general structures we find apply more generally,
to any configuration involving collections of branes intersecting at
angles in more general varieties \footnote{For more general possibilities
involving higher dimensional branes see \cite{bgkl}, and the more
extensive analysis in \cite{afiru}. See also \cite{kachru} for systems of
D6-branes on 3-cycles in general Calabi-Yau spaces.}. Another point worth
mentioning is that the case of D4-branes admits an interesting M-theory
lift. Indeed, D4-branes correspond to M-theory 5-branes wrapping on the
eleventh dimension, compactified on a circle $\IS^1$. Thus the models
discussed in the paper may be regarded as M-theory compactifications on
$\IS^1 \times \IT^2\times {\bf B_4}$ with M5-branes wrapping on
$\IS^1\times \IT^2$.

There are a number of issues to be further studied. On the
theoretical side, the brane configurations we have considered are
non-supersymmetric, and hence the question of their stability
deserves further study. In this regard, it is worth mentioning
that (meta)stable  configurations on analogous models using
wrapping D6-branes have been recently discussed in \cite{afiru}.
Also, consideration of more general string configurations with
branes at angles could lead to improvements in model building in
this setup. On the more phenomenological side, it should be
interesting to carry out a general study of possible
three-generation models leading to interesting gauge coupling
predictions, and fermion mass textures, using the built-in
mechanism for the generation of hierarchies in this class of
models. There are other aspects that we have not discussed, such
as the question of neutrino masses, or the strong CP problem. It
should be interesting to examine whether this scenario provides
some new understanding for these questions. Finally, the study of
signatures of the different KK, windings and gonion particles at
accelerators should also be interesting. Unlike other string
scenarios, this seems to be amenable to direct experimental test.

In summary, we believe that the intersecting brane worlds setup
provides new ways to look at the specific physics of brane world
scenarios  with a low string scale. It also suggests natural
solutions to some of its potential problems, like proton
stability and predicts the presence of new  KK/winding and gonion
particles in between the weak and the string scales which should
be accessible to future colliders.
 It would be interesting to work out in
more detail the predictions of this scenario which could perhaps
provide an exciting alternative
to the much more studied case of low-energy supersymmetry.

\centerline{\bf Acknowledgements}

We thank R.~Blumenhagen, B.~K\"ors, D.~L\"ust, F. ~Marchesano and F.~Quevedo  
for useful discussions. A.~M.~U. thanks the Instituto Balseiro, CNEA,
Centro At\'omico Bariloche, Argentina, for hospitality, and M.~Gonz\'alez
for encouragement and support. G.A work is partially supported by ANPCyT
grant 03-03403. L.E.I. and R.R. are partially supported by CICYT (Spain)
and  the European Commission (grant ERBFMRX-CT96-0045).

\end{document}